\documentclass[aps, prd, twocolumn, nofootinbib, longbibliography, superscriptaddress, preprintnumbers]{revtex4-1}

\usepackage{epsfig}
\usepackage{enumitem}
\usepackage{amsmath}
\usepackage{amssymb}
\usepackage{graphicx}
\usepackage{booktabs}
\usepackage{epstopdf}
\usepackage{gensymb}
\usepackage{aas_macros}
\usepackage{multirow}
\usepackage{lipsum}
\usepackage{fontawesome5} 
\usepackage{physics}
\usepackage{makecell}
\usepackage{enumitem}
\usepackage{comment}
\usepackage{bm}
\usepackage{siunitx}
\usepackage{threeparttable}
\usepackage{array}

\newcommand{\blfootnote}[1]{%
  \begingroup
  \renewcommand{\thefootnote}{}\footnote{#1}%
  \addtocounter{footnote}{-1}%
  \endgroup
}

\usepackage[table,xcdraw,dvipsnames]{xcolor}

\usepackage[dvipsnames,table,xcdraw]{xcolor}  
\definecolor{mycitecolor}{HTML}{ED0DD9}

\usepackage[colorlinks=true,citecolor=teal,linkcolor=teal]{hyperref}
\usepackage[capitalise,noabbrev]{cleveref}

\usepackage{tikz,xcolor,hyperref}

\definecolor{lime}{HTML}{A6CE39}
\DeclareRobustCommand{\orcidicon}{\hspace{-4pt}
	\begin{tikzpicture}
		\draw[lime, fill=lime] (0,0) 
		circle [radius=0.16] 
		node[white] {\hspace{0.1mm}{\fontfamily{qag}\selectfont \tiny ID}};
		\draw[white, fill=white] (-0.07,0.1) 
		circle [radius=0.01];
	\end{tikzpicture}
	\hspace{-3.2mm}
}

\foreach \x in {A, ..., Z}{\expandafter\xdef\csname 
	orcid\x\endcsname{\noexpand\href{https://orcid.org/\csname orcidauthor\x\endcsname}
		{\noexpand\orcidicon}}
}

\def\bi{\begin{itemize}[noitemsep,leftmargin=2em]
\setlength\itemsep{0.1em}
        }
\def\ei{\end{itemize}}
\hypersetup{
    colorlinks = true,
    linkcolor  = teal,
    citecolor  = teal,
    urlcolor   = teal
}

\begin{document}
\flushbottom
\preprint{FERMILAB-PUB-25-0860-T}

\title{Possible $\nu$ Source Class: 3-sigma Detection of High-Energy Neutrinos from Supermassive Black Hole Binary Candidates}
\blfootnote{The first two authors contributed equally to this work.}

\author{Pugazhendhi A D\orcidD{}}
\email{pugazhendhia@iisc.ac.in}
\affiliation{Department of Physics, Indian Institute of Science, C.\,V.\,Raman Avenue, Bengaluru 560012, India}
\affiliation{Centre for High Energy Physics, Indian Institute of Science, C.\,V.\,Raman Avenue, Bengaluru 560012, India}

\author{Subhadip Bouri\orcidA{}}
\email{subhadipb@iisc.ac.in}
\affiliation{Department of Physics, Indian Institute of Science, C.\,V.\,Raman Avenue, Bengaluru 560012, India}
\affiliation{Centre for High Energy Physics, Indian Institute of Science, C.\,V.\,Raman Avenue, Bengaluru 560012, India}

\author{Bei Zhou\orcidB{}}
\email{beizhou@fnal.gov}
\affiliation{Theory Division, Fermi National Accelerator Laboratory, Batavia, Illinois 60510, USA}
\affiliation{Kavli Institute for Cosmological Physics, University of Chicago, Chicago, Illinois 60637, USA} 

\author{Rachana\orcidE{}}
\email{rachana2022@iisc.ac.in}
\affiliation{Joint Astronomy Programme, Department of Physics, Indian Institute of Science, C.\,V.\,Raman Avenue, Bengaluru 560012, India}
\affiliation{Indian Institute of Astrophysics, Block II, Koramangala, Bangaluru 560034, India}
\author{Ranjan Laha\orcidC{}}
\email{ranjanlaha@iisc.ac.in}
\affiliation{Centre for High Energy Physics, Indian Institute of Science, C.\,V.\,Raman Avenue, Bengaluru 560012, India}

\date{\today}

\begin{abstract}
Identifying the sources of high-energy (TeV--PeV) astrophysical neutrinos is crucial for studies in both astrophysics and particle physics. 
Despite extensive searches for more than a decade, which revealed several individual potential sources and only one potential source class, the origins of these neutrinos remain largely unresolved; thus, more source classes should be investigated.
\textit{In this work}, we conduct the first search for high-energy neutrino emission from a new source class, supermassive black hole binaries (SMBHBs), which are also theoretically motivated.
We perform an unbinned maximum-likelihood-ratio analysis on our constructed catalog of 693 SMBHB candidates and 10 years of IceCube public data.
Our results show positive correlations, with higher significance in more physically motivated scenarios and the highest significance at \textbf{{\color{magenta}3.0$\sigma$}}. In addition, we also study potential connections between SMBHBs' high-energy neutrino and nano-Hz gravitational-wave emissions, the latter being the main target of pulsar timing arrays.
Our results provide the first evidence of SMBHBs being high-energy neutrino emitters. 
\end{abstract}
\maketitle
\section{Introduction}

The discovery of high-energy (HE) diffuse astrophysical neutrinos\,\footnote{Throughout the work, we do not distinguish between neutrinos and antineutrinos} by the IceCube Neutrino Observatory has opened a new chapter in the study of astrophysics and particle physics in the most extreme environments of the Universe~\cite{IceCube:2013cdw, IceCube:2015qii, Aartsen:2013jdh, Ahlers:2018fkn, IceCube:2021rpz, Ackermann:2022rqc}.
Subsequent to IceCube, ANTARES and Baikal-GVD Collaborations also reported their detections~\cite{ANTARES:2024ihw, Baikal-GVD:2022fis}.
Earlier this year, the KM3NeT Collaboration reported a neutrino event with the highest energy ($\sim$220~PeV) ever detected~\cite{KM3NeT:2025npi, KM3NeT:2025vut, KM3NeT:2025aps,KM3NeT:2025bxl}. These detections have made HE neutrinos powerful tools for understanding the extreme Universe, establishing neutrinos as an indispensable component of the multimessenger framework; see, e.g., Ref.~\cite{Ackermann:2022rqc} for a recent review.

Identifying the sources of these diffuse neutrinos is a critical step in both astrophysics and particle physics. In astrophysics, this will reveal the origins of HE cosmic rays, a long-standing puzzle, and the nature of their accelerators. 
In particle physics, this will make these neutrinos much sharper tools for testing new physics, i.e., with known directions and cosmic or Galactic baselines. 
For more than a decade, significant efforts have been made to search for these sources, by both experimental collaborations and theorists. 
The first potential source reported by the IceCube Collaboration was TXS 0506+056, a blazar that showed two independent neutrino flares, both with $\simeq3\sigma$ global significance~\cite{IceCube:2018cha, IceCube:2018dnn}.
The second is NGC 1068~\cite{IceCube:2019cia, IceCube:2022der}, a Seyfert galaxy that showed a $\simeq 4.2\sigma$ global significance in time-independent HE neutrino emission.
However, these two sources together can only explain $\sim1\%$ of the total HE astrophysical neutrino flux~\cite{IceCube:2018cha, IceCube:2022der}~\footnote{The Galactic plane was detected as diffuse emission at 4.5$\sigma$~\cite{IceCube:2023ame}.}.

The majority of astrophysical HE neutrinos are expected to be produced by certain source classes.
Recent analyses by the IceCube Collaboration have identified the \textit{first potential source class}---X-ray-bright Seyfert galaxies, such as NGC 1068, NGC 4151, CGCG 420-015, and others~\cite{IceCube:2024dou, IceCube:2024ayt, Abbasi:2025tas}. The most recent search~\cite{Abbasi:2025tas} found a 3.3$\sigma$ excess from an ensemble of 11 such sources (without NGC 1068).
Extensive searches have also investigated other source classes, both galactic and extragalactic, but no significance was found, such as gamma-ray bursts~\cite{IceCube:2009ror, IceCube:2011vle, IceCube:2012qza, IceCube:2014jkq, IceCube:2016ipa, IceCube:2017amx, IceCube:2023woj},
different types of active galactic nuclei (AGN)~\cite{IceCube:2016qvd, IceCube:2021pgw, Zhou_2021, Creque-Sarbinowski:2021nil, Abbasi:2022uox, IceCube:2023htm,
Bellenghi:2023yza, IceCube:2025ggw},
different types of supernovae~\cite{Senno:2017vtd, Esmaili:2018wnv, Chang:2022hqj, IceCube:2023esf, Lu:2025jks},
fast radio bursts~\cite{Fahey:2016czk, IceCube:2017fpg, IceCube:2019acm, Desai:2021dpm, IceCube:2022mjy}, galaxy mergers~\cite{Bouri:2024ctc},
supernova remnants~\cite{IceCube:2023esf,ANTARES:2020srt}, 
pulsar wind nebulae~\cite{IceCube:2020svz}, 
X-ray binaries~\cite{IceCube:2022jpz},
and many others.
Despite these efforts, the origins of HE astrophysical neutrinos \textit{remain largely unresolved}. Thus, \textit{more source classes should be investigated}.

In this work, we conduct the first search for HE neutrinos from supermassive black hole binaries.
At the heart of most galaxies, a supermassive black hole (SMBH), with a typical mass of $10^6 M_\odot$--$10^{10} M_\odot$, is situated. When two galaxies merge over a billion-year timescale, the central SMBHs are drawn to each other by gravity and orbit around a common center of mass, forming an SMBHB. 
Our search is also motivated by recent theoretical-modeling developments for the HE neutrino emission from SMBHBs~\cite{Jaroschewski_2022, Yuan:2020oqg, BeckerTjus:2022leo, deBruijn:2020pky}.
For example, Ref.~\cite{Jaroschewski_2022} demonstrated that inspiraling SMBHB systems can significantly contribute to the IceCube measured diffuse astrophysical neutrino flux. 
Ref.~\cite{Yuan:2020oqg} discussed the jet-induced neutrino emission from SMBHB mergers and showed that the relativistic post-merger jets interacting with the dense circumnuclear wind can produce detectable PeV neutrino fluxes.
Ref.~\cite{BeckerTjus:2022leo} discussed that the observed episodic neutrino flare of TXS 0506+056 can be explained by a precessing jet powered by an SMBBH system in its inspiral phase.

SMBHB systems are also promising gravitational-wave (GW) sources in the nano-Hz frequency band, which is the main target of pulsar timing arrays. Their collective GW signal has recently been reported by several pulsar-timing-array collaborations~\cite{Antoniadis:2023A50, Agazie:2023L8, Reardon:2023L6, Xu:2023wog}.
Moreover, SMBHB coalescences serve as powerful laboratories for testing gravity in the strong-field regime. Thus, detection of neutrinos from SMBHBs would enrich these systems as multimessenger emitters.

We use 10 years of public IceCube data~\cite{website_neutrino_data} together with a catalog of SMBHB candidates we construct based on Ref.~\cite{SMBBHScatalog}.  
We perform a maximum-likelihood-ratio analysis, a standard method for searching for HE neutrino sources. 
The main finding of this work is a \textbf{\textcolor{magenta}{3.0$\sigma$}} correlation in our most-physically-motivated scenario and slightly lower significance in our less-physically-motivated scenarios.
We also investigate potential connections between the detected HE neutrino emission and expected GW signals.

The paper is organized as follows. In Sec.~\ref{sec:nu_data}, we describe IceCube's 10-year public data used in this work. 
In Sec.~\ref{sec_catalog}, we present our constructed catalog of SMBHB candidates.
In Sec.~\ref{sec:formalism}, we discuss the statistical formalism used for our analysis. 
In Sec.~\ref{sec_results}, we present our results. 
In Sec.~\ref{sec:gw_smbhb} we investigate the connections between HE neutrino and GW emissions.
We conclude and discuss the prospects motivated by our findings in Sec.~\ref{sec:conc}.

Throughout this paper, we adopt the cosmological parameters measured by the Planck 2018 mission: $\Omega_M = 0.30966$, $\Omega_\Lambda =  0.69034$, $\Omega_k=0$ and $H_0=67.77$ km/(Mpc. sec)~\cite{Planck:2018vyg}.

\section{IceCube Neutrino Data}\label{sec:nu_data}

The IceCube Neutrino Observatory (hereafter IceCube), located at the Amundsen-Scott South Pole Station in Antarctica, utilizes a cubic kilometer of pristine Antarctic ice to detect neutrinos.
IceCube consists of 5160 digital optical modules (DOMs) suspended on 86 strings deep within the ice at depths ranging from 1,450 to 2,450 meters~\cite{IceCube:2006tjp, IceCube:2008qbc, IceCube:2013dkx, IceCube:2016zyt}. Through these DOMs, IceCube can detect the Cherenkov light emitted by electrically charged secondary particles from all-flavor neutrino interactions. 

Fig.~\ref{fig:source_loc} shows the distribution of the data that we use in our analysis, which is the 10-year public muon-track events released by the IceCube Collaboration~\cite{Abbasi:2021bvk,website_neutrino_data}. The dataset comprises 1,134,450 events spanning from April 6, 2008, to July 8, 2018. 
These events are divided into five sub-datasets, corresponding to five different phases of IceCube construction: (i) IC40, (ii) IC59, (iii) IC79, (iv) IC86-I, and (v) IC86-II to IC86-VII~\cite{2011ApJ...732...18A, IceCube:2013kvf, IceCube:2014vjc, IceCube:2019cia}. 
The numbers associated with the sub-dataset names refer to the total number of strings with DOMs that were deployed during that particular phase. 
The dataset was later revised by Ref.~\cite{Zhou:2021xuh} and made publicly available through a \faGithub\ \href{https://github.com/beizhouphys/IceCube_data_2008--2018_double_counting_corrected}{\textcolor{teal}{Github Repository}}. 
The revision was to correct the $19 \times 2$ double-counted events in the dataset found in Ref.~\cite{Zhou:2021xuh} (listed in its table III). 
These events arise from a previous internal reconstruction error (now corrected) that misidentified some single muons crossing the dust layer as two separate muons arriving at the same time and in close direction~\cite{Zhou:2021xuh}. 
We use this corrected dataset for our analysis. We consider events with declination between $-10^\circ$ and $90^\circ$, excluding Dec~$<-10^\circ$ due to high atmospheric muon backgrounds.

The muon tracks mainly come from the neutrino-nucleus deep-inelastic scattering~\cite{ParticleDataGroup:2024cfk},
but also from neutrino-nucleus W-boson production~\cite{Seckel:1997kk, Alikhanov:2015kla, Zhou:2019vxt, Zhou:2019frk, Xie:2023qbn}. 
Furthermore, Ref.~\cite{Plestid:2024bva} pointed out the importance of final-state radiation, which lowers the energy of the muon tracks by up to 5\% in this dataset. The contribution from the Glashow resonance~\cite{Glashow:1960zz, IceCube:2021rpz} is minor, as it is around 6.3~PeV, where the neutrino flux is low.

\section{Catalog of SMBHB candidates}
\label{sec_catalog}

Although no SMBHBs have been individually detected via GW or electromagnetic observations, the latter have identified hundreds of SMBHB candidates. 
Fig.~\ref{fig:source_loc} shows the distribution of SMBHB candidates in the catalog we construct based on Ref.~\cite{SMBBHScatalog}.
These candidates are identified through various electromagnetic observations, including spectral signatures, photometric signatures for periodicity, and jet morphology, detailed in Sec.~\ref{sec_catalog_signatures}. By combining candidates from multiple observational channels and refining the list based on available follow-up studies, we construct a catalog of \textbf{693} SMBHB candidates. Among these, \textbf{335} are classified as Type-I quasi-stellar objects (QSOs), for which bolometric luminosities and Eddington ratios are provided and utilized in some of our analysis.
This catalog, derived primarily from optical surveys in the northern hemisphere, exhibits three distinct source concentrations in the northern sky, with voids along the Galactic plane where foreground contamination hinders reliable extragalactic identification. 
  
Several studies have theoretically modeled neutrino emission from SMBHB systems~\cite{Jaroschewski_2022, Yuan:2020oqg}.
In this study, we follow Ref.~\cite{Jaroschewski_2022}, which discusses the SMBHB merger in four different stages: 
(1) dynamical friction stage~\cite{Gold:2019nqg, Yu:2001xp}, 
(2) final parsec stage~\cite{Jaffe:2002rt, Volonteri:2002vz, Gold:2019nqg, Volonteri:2002vz, Begelman:1980vb, Zier:2002gr}, 
(3) inspiral stage~\cite{Gergely:2007ny, Blandford:1982xxl, Blandford:1977ds, Daly:2019srb, Kun:2018dqx}, 
and (4) final merger stage.
Of these, the inspiral stage is of primary interest to us. 
The main dissipative effect in this stage is the gravitational radiation~\cite{Gergely:2007ny, Blandford:1982xxl, Blandford:1977ds}. 
In addition, the unaligned spins of the SMBHs precess and gradually realign with the orbital angular momentum, leading to a spin-flip of the jets~\cite{Daly:2019srb}. 
During the SMBH spin realignment, the jet changes its direction and sweeps through the surrounding medium. This causes collisions of HE protons, which produce pions: $p + p \rightarrow N_{\pi}\, [\,\pi^{0} + \pi^{+} + \pi^{-} ] + X $, where $N_{\,\pi}$ is the pion multiplicity and X are the other hadrons. The produced pions decay into two neutrino flavors via $\pi^{+}\,(\pi^{-}) \rightarrow \mu^{+}\,(\mu^{-}) + \nu_{\mu}\,(\bar{\nu}_{\mu}),$ and $\mu^{+}\,(\mu^{-}) \rightarrow e^{+}\,(e^{-}) + \nu_{e}\,(\bar{\nu}_{e}) + \bar{\nu}_{\mu}\,(\nu_{\mu})$. These neutrinos could be detected by neutrino observatories.

\subsection{Catalog preparation for our analysis}
\label{sec:catalog}

Apart from the two HE neutrino hotspots mentioned in the introduction, TXS 0506+056 and NGC 1068, the IceCube Collaboration has identified several other neutrino hotspots, potentially from extragalactic sources, as follows. Refs.~\cite{MAGIC:2025tlp, IceCube:2024ayt} identified NGC~4151 with a post-trial significance of  $\sim 2.9\sigma$.
Ref.~\cite{IceCube:2024dou} obtained a global significance of $\sim 2.7\sigma$ from a combined analysis of NGC~4151 and CGCG~420-015. In addition, Ref.~\cite{IceCube:2022der} identified PKS 1424+240.
Moreover, recent IceCube analyses~\cite{Halzen2025_LHAASO, Abbasi:2025tas} reported a binomial-excess signal driven by several sources. In particular, NGC~7469 (local significance: 3.8$\sigma$), MCG~4-48-2 (1.7$\sigma$), Cygnus~A (2.5$\sigma$), Mrk~1498 (1.7$\sigma$), NGC~3079 (1.7$\sigma$), NGC~4992 (2.0$\sigma$), Mrk~417 (1.6$\sigma$), LEDA~166445 (2.1$\sigma$), CGCG~420-015 (2.7$\sigma$), and NGC~1194 (1.8$\sigma$) contribute to the observed excess, making them potential candidates for HE neutrino emission. Our catalog has NGC 4151 and TXS 0506+056.

Therefore, we prepare \textit{three nested catalogs} for our analysis, as follows.

\textbf{(1) Whole catalog}: our default catalog.

\textbf{(2) Association-free catalog}: excluding eight sources associated with IceCube-detected hotspots, as follows. 
TXS 0506+056, NGC 4151, four sources close to the NGC 1068 hotspot (SDSS J024442.77-004223.2, J024613.89-004028.2, J024703.24-010032, and J024455.18-002501.5), and two sources located near NGC 1194 and Mrk 417~\cite{IceCube:2024ayt}.

This approach ensures that our results are not biased by previously identified candidates.

\textbf{(3) NANOGrav catalog:} NANOGrav Collaboration~\cite{Agarwal:2025cag} has conducted the first targeted search for continuous GW signals from 114 SMBHB candidates with their 15-year dataset. From this sample, we select 104 sources in the northern hemisphere, which are a subset of our ``whole/association-free catalogs''. These systems exhibit strong multiwavelength evidence supporting their classification as SMBHB candidates.

\subsection{SMBHB candidates in our catalog}
\label{sec_catalog_signatures}

In the following subsubsections, we discuss the methods from electromagnetic observations used to identify the SMBHB candidates in our catalog. 

\subsubsection{Spectral signatures} 

In active galactic nuclei (AGN), two different types of spectral lines from gas emission are observed: broad lines, produced by Doppler broadening of the high-velocity gas near the SMBH,
and narrow lines, originating from low-velocity gas located farther from the SMBH.
Often, if a significant offset between these lines is observed, it may be a clue for the SMBHB system. 
The light from the gas close to one of the SMBHs is redshifted or blueshifted due to its orbital motion, causing the observed offset. 
Numerous studies have largely used this technique to search for binary systems. We include sources in our catalog from the studies referenced in Ref.~\cite{SMBBHScatalog}, as follows.

\begin{itemize}[leftmargin=1.0em, itemsep=2pt, topsep=1pt, parsep=0pt, partopsep=0pt]

\item Tsalmantza et al.~(2011)~\cite{Tsalmantza_2011}: searches for broad line offsets with respect to narrow lines in quasars with $0.1 < z < 1.5$. We include 8 potential candidates from this study in our catalog.

\item Eracleous et al.~(2012)~\cite{Eracleous_2012}, 
Shen et al. (2013)~\cite{Shen_2013}, and Liu et al. (2014b)~\cite{2014ApJ...789..140L}: searches for shifts in the broad $H\beta$ lines from the rest frame of quasars with $z < 0.7$. 
We include 29 candidates from these studies in our catalog.

\item Ju et al.~(2013)~\cite{Ju_2013}: searches for radial velocity shifts in normal, single-peaked broad MgII emission lines in quasars with $0.36 < z < 2$. We include 48 candidates from this study in our catalog.

\item From individual SMBHB candidates identified primarily in early SDSS studies, as mentioned in Ref.~\cite{SMBBHScatalog}, we add two sources to our catalog. 
\end{itemize}

\subsubsection{Photometric signatures} 

The photometric technique is based on the periodic variability in the observed light curves. 
The orbital motion of a binary system imprints this periodic variability through different physical processes. Several types of variability are observable, including hydrodynamical variability, orbital Doppler boost, and binary self-lensing. 
Several studies have applied these techniques to detect periodicity in quasar light curves, and those that we have incorporated into our catalog are as follows:

\begin{itemize}[leftmargin=1.0em, itemsep=2pt, topsep=1pt, parsep=0pt, partopsep=0pt]

\item Graham et al.~(2015b)~\cite{2015MNRAS.453.1562G}: analyzed 243,500 spectroscopically confirmed quasars from the Catalina Real-Time Transient Survey to search for strong Keplerian periodic signals exhibiting at least 1.5 cycles over a nine-year baseline. A total of 109 candidates identified in this study are included in our catalog.

\item Charisi et al.~(2016)~\cite{2016MNRAS.463.2145C}: conducted a statistical search for periodic optical brightness variability in 35,383 quasars from the Palomar Transient Factory using Lomb–Scargle periodograms, assessing significance via a damped random walk model. We include 50 candidates from this study in our catalog.

\item Liu et al.~(2019~\cite{2019ApJ...884...36L}: investigation of 9,000 quasars from Pan-STARRS Medium Deep Survey, exhibiting periodic brightness variations over multiple cycles. We include one candidate from this study in our catalog.

\item Chen et al.~(2020)~\cite{DES:2020lsu}: Analysed 625 spectroscopically confirmed quasars in DES+SDSS Stripe 82 with 20-year multi-color light curves, identifying five potential periodically variable quasar candidates with observed periods of ~35 years (12 years in rest frame). We include these five candidates in our catalog.

\item Chen et al.~(2022b)~\cite{chen2023searchingquasarcandidatesperiodic}: Searches 143,700 quasars in ZTF, identifying quasars with multi-cycle periodic or quasi-periodic variations validated against red-noise models; we include 127 potential candidates from this study into our catalog.

\item Apart from these, several periodic candidates were identified individually, as mentioned in Ref.~\cite{SMBBHScatalog}, from which we include 22 potential sources in our catalog.
\end{itemize}

\subsubsection{Jet morphology}

A way to recognize SMBHB is to observe its jet morphology, which often exhibits ``winged'' or X-shaped radio structures.
These morphologies may arise from misaligned jets associated with each black hole in a binary system or from a spin-flip of the central SMBH triggered by a recent merger. Ref.~\cite{2019ApJS..245...17Y} provides a catalog of 290 radio galaxies exhibiting the winged and X-shaped jet morphologies. 
We include these sources in our catalog.

\begin{figure*}[t!]
\centering
\includegraphics[scale=0.25]{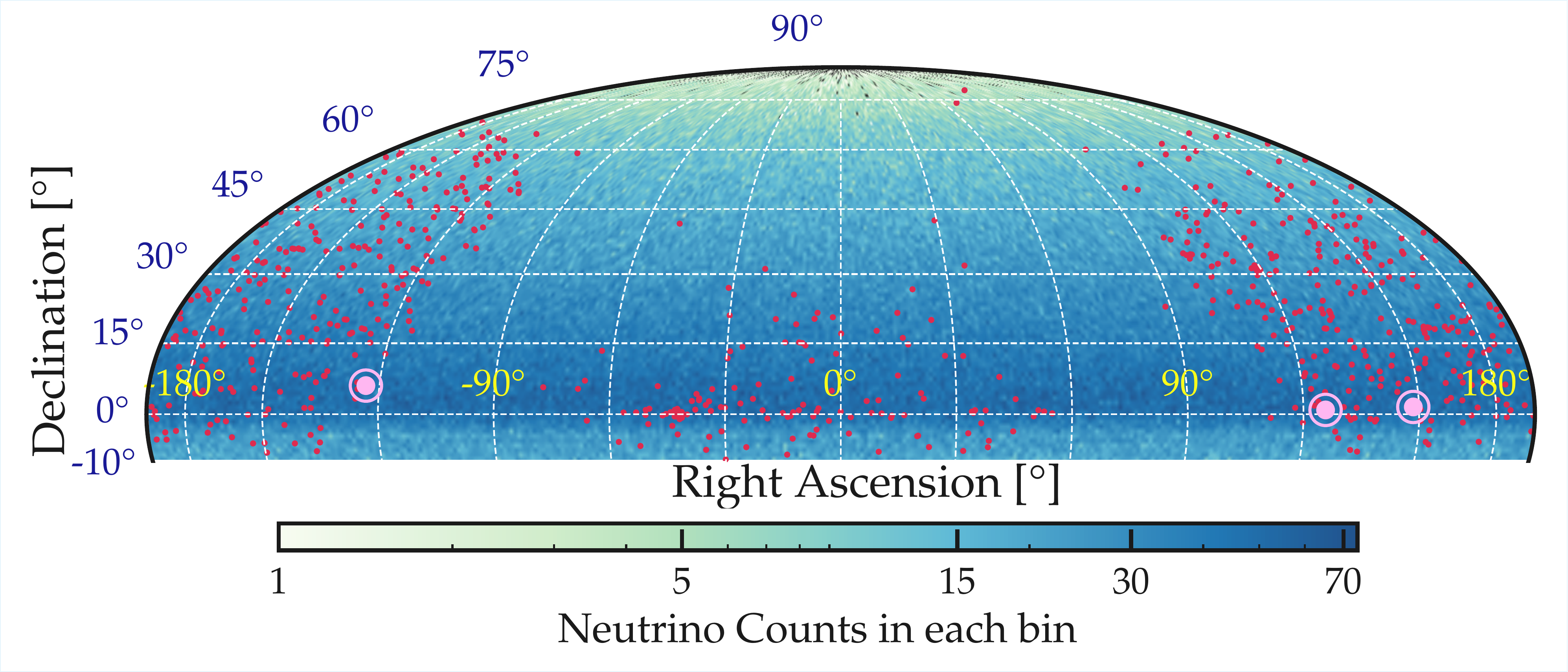}
\caption{
Sky map in the equatorial coordinate system with the neutrino events and sources from our catalog.
The \textcolor[HTML]{31688E}{\textbf{blue gradient}} shows the \textcolor[HTML]{31688E}{\textbf{distribution of events}} from the 10-year IceCube public data (April 2008--July 2018). The sky is divided into $1^{\circ} \times 1^{\circ}$ bins, with the color indicating the number of detected neutrino events in each bin.
The \textcolor{red}{\textbf{red dots}} indicate the \textcolor{red}{\textbf{locations of the SMBHB candidates}} from the catalog we construct, described in Sec.~\ref{sec_catalog}, while the \textcolor[HTML]{CC00CC}{\textbf{ encircled pink markers}} represent the \textcolor[HTML]{CC00CC}{three notable sources that contributed most} to the 3$\sigma$ found in our work (see Sec.~\ref{sec_results}). 
}
\label{fig:source_loc} 
\end{figure*}

\section{Formalism}\label{sec:formalism}
\label{sec_formalism}

We use the unbinned maximum-likelihood-ratio method to search for neutrino signals from the SMBHB candidates in our catalog~\cite{Braun_2010, Braun_2008, Braun:2008bg, BRAUN2010175}. 
This method tests the signal-plus-background hypothesis against the background-only (or null) hypothesis. It is a well-established technique widely used in HE neutrino astrophysics by the IceCube (e.g., Refs.~\cite{IceCube:2010nca, IceCube:2013kvf}), KM3NeT (e.g., Refs.~\cite{Benfenati:2023uuz, KM3NeT:2024uhg}) and ANTARES (e.g., Refs.~\cite{ANTARES:2020zng, ANTARES:2017pax}) Collaborations, as well as in independent studies outside these collaborations (e.g., Refs.~\cite{Hooper:2018wyk, Zhou_2021, Chang:2022hqj,Bouri:2024ctc, Li:2022vsb, M:2025oyf}).

The likelihood function is constructed as the product of the likelihood for each neutrino event, indexed by $i$, within a sub-dataset, indexed by $k$:
\begin{equation}
\mathcal{L}(n_s, \Gamma) = \prod_k \prod_{i \in k} \left[ \frac{n_s^k}{N_k} S_i^k + \left( 1 - \frac{n_s^k}{N_k} \right) B_i^k \right],
\label{eq_LLH}
\end{equation}
where $S_i^k$ and $B_i^k$ are the signal and background probability density functions (PDFs).
The parameters $n_s$ and $\Gamma$ denote the total number of signal-like events and the neutrino spectral index, respectively, for the whole dataset. Both parameters are treated as free and are determined by maximizing the likelihood. 
The number of signal events in each sub-dataset, $n_s^k$, is related to $n_s$ by
$n_s^k = f_k \, n_s$,
where $f_k$ represents the expected fractional contribution of the $k^{\mathrm{th}}$ sub-dataset to the total signal, given an assumed spectral index $\Gamma$. 
The explicit expressions for $f_k$ are provided below.
Further details of the analysis framework can be found in Refs.~\cite{Zhou_2021, Chang:2022hqj, Bouri:2024ctc}.

The test statistic (TS) is defined to measure how strongly the IceCube data favor the signal-plus-background hypothesis over the background-only hypothesis:
\begin{equation}
{\rm TS}(\hat{n}_s, \hat{\Gamma})=2\,\ln \frac{\mathcal{L}(\hat{n}_s,\hat{\Gamma})}{\mathcal{L}(n_s=0)} \,,
\end{equation}
where $\hat{n}_s$ and $\hat{\Gamma}$ are the values of $n_s$ and $\Gamma$ that maximize the likelihood in Eq.~\eqref{eq_LLH}. The denominator $\mathcal{L}(n_s=0)$ corresponds to the background-only hypothesis, i.e., all observed events come from the background. A high TS value reflects the presence of neutrino events coming from the sources.

The statistical significance of a potential signal, as well as the upper limits in the absence of a detection, can be evaluated as follows.
In our case with large samples of data, according to Wilks’ theorem~\cite{Wilks:1938dza}, the TS approximately follows a $\chi^2$ distribution with a number of degrees of freedom equal to the number of free parameters in the analysis. 
The observed TS value can thus be converted to the corresponding $p$-value under the $\chi^2$ distribution, from which the significance can be obtained in units of standard normal deviations.
\textit{Throughout the paper, all the upper limits (ULs; on the $n_s$, hence on the flux) are at 95\% confidence level (CL)}, which can be obtained from $\rm -2\Delta\rm{ln}\mathcal{L}=TS-3.84$ and 5.99 for the analysis with one and two free parameters, respectively.

The following equation connects $n_s$, the expected number of $\nu_{\mu} + \bar{\nu}_\mu$ events at IceCube, with the neutrino fluxes,
\begin{equation}
n_s = 2\pi \sum_k t_k \int d\sin\delta \int A_{\text{eff}}^k(E_{\nu}, \delta_j) \,\frac{dF}{dE_\nu} dE_{\nu}\,,
\label{eq:expected_ns}
\end{equation}
where $t_k$ and $A_{\text{eff}}^k(E_{\nu}, \delta_j)$ are the detector's uptime and effective area, respectively, in the $k$-th sub-dataset, $E_\nu$ is the neutrino energy, and $\delta$ is the declination angle. 
Throughout this paper, we use the per-flavor neutrino flux/spectrum (e.g., $\nu_\mu+\bar{\nu}_\mu$) and assume that the neutrino spectrum from the sources follows a single-power-law with spectral index $\Gamma$, i.e.,
\begin{equation}
\dd F_\nu(E_{\nu})/\dd E_{\nu} = \phi_0 \left( E_{\nu}/100~\text{TeV} \right)^{\Gamma}\,.
\label{eq_flux}
\end{equation}

\subsection{Stacking analysis}
\label{sec_formalism_stacking}

As this paper primarily focuses on neutrino emission from the SMBHB source class, we conduct a stacking analysis to search for neutrino signals from the catalogs.
The stacking analysis are more powerful than the single-source analysis, which looks for neutrino emission from individual sources in the catalogs.

We use the spatial and energy terms of the signal and background PDFs, as in Refs.~\cite{Zhou_2021, Chang:2022hqj, Bouri:2024ctc} and many IceCube Collaboration papers.
The signal PDF, $S_{i}^k$, is given by \begin{equation}
S_{i}^k = \frac{\sum_j w_{j, \rm model}\,w_{j, {\rm acc}}^k \,S_{{\rm spt},\,ij}^k}{\sum_j w_{j, \rm model}\,w_{j, {\rm acc}}^k}\,S_{E,\,i}^{k},
\end{equation} 
where $w_{j, {\rm acc}}^k$ is the weighting factor that depends on the detector's response and is given by
\begin{equation}
w_{j, {\rm acc}}^k(\delta_j) \propto t_k \times \int A_{\rm eff}^k(E_{\nu},\delta_j)E_{\nu}^{\Gamma}dE_{\nu} \,,
\label{eq:weighting_fac}
\end{equation}
where $A_{\rm eff}^k(E_{\nu},\delta_j)$ is the same as that defined in Eq.~\eqref{eq:expected_ns},  and $w_{j, \rm model}$ is the source-property-dependent weighting factor discussed below. For the signal PDF, $S_{{\rm spt},\,ij}^k$ is the spatial-signal PDF defined as,
\begin{equation}
S_{{\rm spt},\,ij}^{k}(\Vec{x_i},\sigma_i,\Vec{x_j})
=
\frac{1}{2\pi\sigma_i^2} \exp\bigg(-\frac{D(\Vec{x_i},\Vec{x_j})^2}{2\sigma_i^2}\bigg) \,,
\label{eq:spatial_signal}
\end{equation}
where $\Vec{x}_i$ and $\Vec{x}_j$ are the directions of the event $i$ and a source indexed by $j$, respectively, and $D(\Vec{x_i},\Vec{x_j})$ is their angular distance.
The $\sigma_i$ is the angular error of the direction of the event. The signal energy PDF, $S_{E,i}^k$, is  given by~\cite{Chang:2022hqj}, 
\begin{multline}
S^k_{E,i}\left(E^{\mathrm{prx}}_i | \delta_i\right)
\propto \int dE_\nu \, E_\nu^{\Gamma} \\
\times A_{\mathrm{eff}}^k(E_\nu, \delta_i)\,
P^k\!\left(E^{\mathrm{prx}}_i | E_\nu, \delta_i\right),
\label{eq:signal_energy_full}
\end{multline}
where $P^k(E^{\mathrm{prx}}_i | E_\nu, \delta_i)$ is the probability 
that a neutrino with true energy $E_\nu$ and declination $\delta_i$ is reconstructed with proxy muon energy $E^{\mathrm{prx}}_i$.

We define the background PDF as $B_i^k=B_{{\rm spt}, i}^k \times B_{E, i}^k$.
The background spatial PDF, $B_i^k=B_{{\rm spt}, i}^k$, is a function of declination $\delta_i$, with an almost isotropic distribution in right ascension due to IceCube's South Pole location. We compute it using the sliding-window method of Ref.~\cite{Zhou_2021}, normalized over $\delta \in [-10^\circ, 90^\circ]$ (see Fig.~2 of Ref.~\cite{Zhou_2021}).
The background energy PDF, $B^k_{E,ij}(E^{\mathrm{prx}}_i | \delta_i)$, 
is defined as the normalized distribution of the reconstructed energy 
proxy $E^{\mathrm{prx}}_i$ as a function of the event declination $\delta_i$. 
It is constructed using a sliding-window method with a window size of 
$\sin\delta_i \pm 0.05$ in declination and $\log_{10}(E^{\mathrm{prx}}_i) \pm 0.2$ 
in reconstructed energy.

The $f_k$ can then be evaluated by 
\begin{equation}
f_k =
\frac{\sum_j w_{j, \rm model}w_{j, {\rm acc}}^k }{\sum_k\sum_j w_{j, \rm model}w_{j, {\rm acc}}^k}.
\label{eq_fk}
\end{equation}

Finally, for $w_{j, \rm model}$, we consider three different weighting schemes, as follows, from the most physically motivated to the least, i.e., model-independent:

\begin{table*}[htbp]
\centering
\renewcommand{\arraystretch}{1.25}
\setlength{\tabcolsep}{6pt}
\textbf{\large Spatial-focused analysis}\\[4pt]
\begin{tabular}{p{2.6cm}|p{2cm}|p{1.5cm}|p{1.6cm}|p{1.6cm}|
>{\centering\arraybackslash}p{1.6cm}|>{\centering\arraybackslash}p{1.6cm}|p{1.1cm}}
\hline
\hline
{Weighting scheme} & {Catalog} & {Number of sources} & {$\hat{\Gamma}$, $\hat{n}_s$} & {TS, sig.} 
& \multicolumn{2}{c|}{\begin{tabular}{@{}c@{}} Flux at 100 TeV \\  $\rm [GeV^{-1}cm^{-2}s^{-1}sr^{-1}]$ \end{tabular}} & {Figure} \\
\cline{6-7}
& & & & & {Best-fit} & {95\% UL} & \\
\hline
\multirow{3}{*}{Neutrino flux} 
 & Whole & 335 & --, 116.3 & 9.09, \textbf{{\color{magenta}3.0$\sigma$}} & 1.10$\times10^{-19}$ & 1.78$\times10^{-19}$ &  \\
\cline{2-7}
 & Assoc.-free & 331 & --, 95.2 & 8.35, \textbf{{\color{magenta}2.9$\sigma$}} & 9.10$\times10^{-20}$ & 1.64$\times10^{-19}$ & Fig.~\ref{fig:neutrino_flux_plot} \\
 \cline{2-7}
 & NANOGrav & 94 & --, 55.8 & 7.59, \textbf{{\color{magenta}2.8$\sigma$}} & 5.30$\times10^{-20}$ & 9.42$\times10^{-20}$ &  \\
\hline

\multirow{3}{*}{Bolometric flux} 
 & Whole  & 335 & --, 136 & 2.62, 1.6$\sigma$ & 1.54$\times10^{-19}$ & 3.44$\times10^{-19}$ &  \\
\cline{2-7}
 & Assoc.-free  & 331 & --, 83.3 & 2.54, 1.6$\sigma$ & 7.92$\times10^{-20}$ & 1.92$\times10^{-19}$ & Fig.~\ref{fig:lbol/dl^2_plot} \\
 \cline{2-7}
 & NANOGrav  & 94 & --& 0.00, 0 & -- & 1.06$\times 10^{-19}$ &  \\
\hline

\multirow{3}{*}{Uniform} 
 & Whole  & 693 & --, 400.2& 2.48, 1.6$\sigma$ &  3.76$\times$10$^{-19}$ &  7.69$\times$10$^{-19}$ &  \\
\cline{2-7}
 & Assoc.-free  & 685 &--, 176.6 & 0.56, 0.7$\sigma$ &  1.66$\times$10$^{-19}$ & 5.97$\times$10$^{-19}$ & Fig.~\ref{fig:uniform_plot} \\
 \cline{2-7}
 & NANOGrav  & 104 & --, 100.5  & 1.00, 1.0$\sigma$ & 4.76$\times10^{-20}$ & 1.50$\times 10^{-19}$ &  \\
\hline
\hline

\end{tabular}

\renewcommand{\arraystretch}{1.1} \centering 
\vspace{0.5cm}

\textbf{\large Spatial-energy analysis}\\[4pt]
\begin{tabular}{p{2.6cm}|p{2cm}|p{1.5cm}|p{1.6cm}|p{1.6cm}|
>{\centering\arraybackslash}p{1.6cm}|>{\centering\arraybackslash}p{1.6cm}|p{1.1cm}}
\hline
\hline
{Weighting scheme} & {Catalog} & { Number of sources} & {$\hat{\Gamma}$, $\hat{n}_s$} & {TS, sig.} 
& \multicolumn{2}{c|}{\begin{tabular}{@{}c@{}} Flux at 100 TeV \\  $\rm [GeV^{-1}cm^{-2}s^{-1}sr^{-1}]$ \end{tabular}}
 & {Figure} \\
\cline{6-7}
& & & & & {Best-fit} & {95\% UL} & \\
\hline
\multirow{3}{*}{Neutrino flux} 
 & Whole & 335 & -3.5, 135.2 & 10.3, \textbf{{\color{magenta}2.5$\sigma$}} & 7.16$\times10^{-22}$ & 1.20$\times 10^{-21}$ &  \\
\cline{2-7}
 & Assoc.-free  & 331 & -3.35, 121.6 & 9.28, 2.3$\sigma$ & 1.39$\times10^{-21}$ & 2.35$\times 10^{-21}$ & Fig.~\ref{fig:neutrino_flux_plot} \\
\cline{2-7}
 & NANOGrav & 94 & -3.6, 68.07 & 7.78, 2.0$\sigma$ & 2.22$\times10^{-22}$ & 3.87$\times 10^{-22}$ &  \\
\hline
\multirow{3}{*}{Bolometric flux} 
 & Whole & 335 & -3.6, 101.6 & 1.66, 0.2$\sigma$ & 3.32$\times10^{-22}$ & 8.51$\times 10^{-22}$ &  \\
\cline{2-7}
 & Assoc.-free & 331 & -3.6, 88.9 & 1.59, 0.1$\sigma$ & 2.91$\times10^{-22}$ & 8.42$\times 10^{-22}$ & Fig.~\ref{fig:lbol/dl^2_plot} \\
\cline{2-7}
 & NANOGrav & 94 & All $\Gamma$ & 0.00, 0 & -- & 1.17$\times 10^{-20}$ &  \\
\hline
\multirow{3}{*}{Uniform} 
 & Whole & 693 & -4.0, 250.2 & 9.19, 2.3$\sigma$ & 1.03$\times10^{-22}$ &1.73$\times 10^{-22}$ &  \\
\cline{2-7}
 & Assoc.-free & 685 &-4.0, 571.1 & 6.30, 1.7$\sigma$ & 2.35$\times 10^{-22}$ & 4.45$\times 10^{-22}$ & Fig.~\ref{fig:uniform_plot} \\
\cline{2-7}
 & NANOGrav & 104 & -3.8, 157.9 & 2.98, 0.7$\sigma$ & 1.93$\times10^{-22}$ & 5.14$\times 10^{-22}$ &  \\
\hline
\hline

\end{tabular}

\caption{
Summary of the key results from our analysis for the three nested catalogs, three weighting schemes, and for both spatial-focused and spatial-energy analysis (Sec.~\ref{sec_results}).}
\label{tab_results}
\end{table*}

(1) \textbf{Neutrino-flux weighting}:
\vspace{-0.03cm}
\begin{equation}
w_{j, \rm model} \propto \frac{f(z)}{(1+z)^2E(z)} \frac{\dd E_{\rm GW}}{dt} g(L) \,,
\end{equation}
motivated by Ref.~\cite{Jaroschewski_2022} (see its Eq.~(2.11), arXiv version).
Here, $f(z)$ and $g(L)$ are the redshift- and luminosity-dependent parts of the luminosity function, respectively. We adopt $f(z)$ as its lower limit for conservativeness. 
$E(z) = \sqrt{\Omega_M(1+z)^3+\Omega_k(1+z)^2+\Omega_\Lambda}$ is the dimensionless Hubble parameter. 
The quantity $\dd E_{\rm GW}/{\dd t}$ is the rate of radiated GW energy emission. Using Eqs.~(2.24) and (2.25) from Ref.~\cite{Jaroschewski_2022} (arXiv version), it is given by 
\begin{align}
\frac{\dd E_{\rm GW}}{\dd t} g(L) \propto
    & \,\frac{\dd (M c^2)}{\dd t} \frac{1}{M} \notag 
\\
& \times \left(\frac{M}{M_{\star}}\right)^{-\alpha}
\exp\left[
  -\left(\frac{M}{M_{\star}}\right)^{\beta}
\right].
\label{eq:luminosity_part}
\end{align} 
In the above equation, 
$\dd (M c^2)/\dd t \propto L_{\text{bol}}$, the Bolometric luminosity, and 
$M \propto L_{\text{Edd}}$, the Eddington luminosity of the sources.  
For further details and the constant values, readers are referred to Sec.~2 in Ref.~\cite{Jaroschewski_2022}. 
Note that $L_{\rm bol}$ is only available for the 335 QSO Type I sources in our catalog, and we apply this weighting scheme to these sources. This physically motivated scheme ensures that the sources with higher expected neutrino flux (nearby or with higher neutrino luminosity) contribute more weight to the likelihood than those with lower expected flux.

(2) \textbf{Bolometric-flux weighting}: 
$w_{j, \rm model} \propto {L_{\rm bol}}/{d_L(z)^2}$,
where \( L_{\rm bol} \) is the bolometric luminosity of the sources. We employ this weighting scheme for the 335  Type-I QSO sources mentioned above.

(3) \textbf{Uniform weighting}: $w_{j, \rm model}=1$, assuming all the SMBHBs have the same HE neutrino flux.

\section{Results from neutrino analysis}
\label{sec_results}

We present the results from two complementary stacking analyses: \textbf{spatial-focused analysis} and a \textbf{spatial-energy analysis}.
The former, which does not include energy PDFs (i.e., no need to assume a spectrum in the analysis), focuses on the spatial clustering correlation between neutrinos and the sources in our catalog. Thus, the analysis has only one free parameter, $n_s$.
Note that a value of $\Gamma$ is still required to compute $f_k$ in Eq.~\eqref{eq_fk}, and we adopt the canonical value of $-2$, but the analysis results depend only very weakly on $\Gamma$.
The latter includes the energy PDFs, presented in Sec.~\ref{sec_formalism_stacking}, and investigates both spatial and spectral correlations. Thus, the analysis has two free parameters, $n_s$ and $\Gamma$.
Adding spectral information could increase the correlation, but if the assumed and underlying spectra are very different, signals could be missed and significance could be reduced.

We vary the spectral index $\Gamma$ from -2 to -4, which encompasses a broad physically and observationally motivated neutrino spectrum~\cite{Aartsen:2020aqd, Abbasi:2021qfz, IceCube:2020acn, IceCube:2024fxo, IceCube:2025ary, IceCube:2015qii, IceCube:2022der, Abbasi:2020jmh}. The maximum value ($\Gamma=-2$) corresponds to the expectation from Fermi 
acceleration mechanisms, while the minimum value ($\Gamma=-4$) is chosen to account for softer spectra that could arise from energy losses, which are stronger at higher energies~\footnote{
A recent analysis from the IceCube Collaboration reported that the astrophysical neutrino flux may be better fit by a broken-power law~\cite{IceCube:2025ewu,IceCube:2025tgp}. 
Our conclusions are robust if we assume a broken-power-law spectrum; see Appendix~\ref{sec:appendix_bpl}.
}.

Table~\ref{tab_results} summarizes the key results from our stacking analysis, with the most significant cases highlighted. The most prominent detection has 3$\sigma$ significance, and several scenarios have 2--3$\sigma$ significance.

Figures~\ref{fig:neutrino_flux_plot}--\ref{fig:uniform_plot} show the results from our stacking analysis for the three nested catalogs, three weighting schemes, and for both spatial-focused and spatial-energy analyses.
In each figure, the left panel shows the variation of $-2\Delta \rm{ln}\,\mathcal{L}$ as a function of $E_\nu^2 dF/dE_\nu$ at $E_\nu = 100$~TeV. 
In the right panel, we show the best-fit energy flux as a function of neutrino energy for cases with $\geq 2\sigma$ significance; for cases with $< 2\sigma$, we present the flux UL at 95\% CL.

Below, we discuss the results of our stacking analysis for each weight scheme.

\subsection{Neutrino-flux weighting scheme}
\label{sec_results_nuflux}

\begin{figure*}[htbp]
\centering
\includegraphics[width=0.95\textwidth]{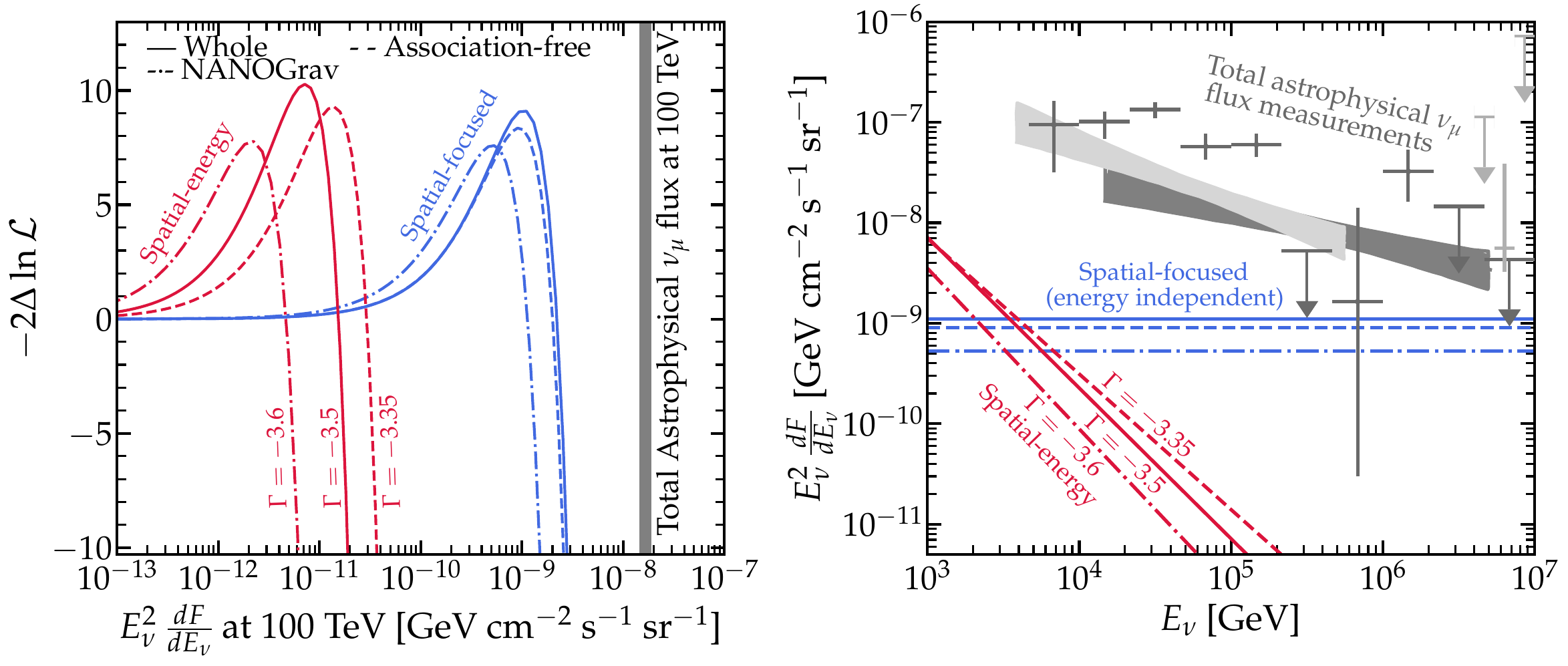}
\caption{
Our stacking-analysis results using the \textbf{neutrino-flux weighting scheme} for the three nested catalogs. The \textbf{\textcolor[HTML]{DC143C}{red curves}} denote the per-flavor neutrino fluxes that maximize the full spatial-energy likelihood, while the \textbf{\textcolor[HTML]{4159E1}{blue curves}} correspond to the spatial-focused likelihood.
\textbf{Left:} log-likelihood values as a function of $E_\nu^2 dF/dE_\nu$ at 100~TeV.
The gray band (\textcolor[HTML]{808080}{$\blacksquare$}) represents the measurements by the IceCube Collaboration from muon track analysis~\cite{IceCube:2021uhz}. 
\textbf{Right:} our inferred neutrino spectra or the UL from the likelihood analysis. 
For the spatial-energy analysis results, the spectral shapes correspond to the best-fit $\Gamma$ from the analysis, whereas for the spatial-focused analysis results, which are energy-independent, we use the canonical $\Gamma=-2$.
For comparison, the grey bands and points with error bars represent the total diffuse HE astrophysical per-flavor neutrino fluxes measured by the IceCube Collaboration using different datasets.
The bands in dark gray (\textcolor[HTML]{808080}{$\blacksquare$}) and light gray (\textcolor[HTML]{D6D6D6}{$\blacksquare$}) were measured using muon-neutrino and starting track events, respectively~\cite{IceCube:2021uhz,Abbasi:2024jro}. 
The dark gray data points with error bars ({\Large \textcolor[HTML]{808080}{\bm{$+$}}}) were measured using combined electron and tau neutrino events~\cite{Aartsen:2020aqd}. The light gray points with errorbars ({\textcolor[HTML]{D6D6D6}{\bm{$+$}}}) were measured using the PeV energy partially contained events~\cite{IceCube:2021rpz}.}
\label{fig:neutrino_flux_plot}
\end{figure*}

\begin{figure*}[htbp!]
\centering
\includegraphics[width=1\textwidth]{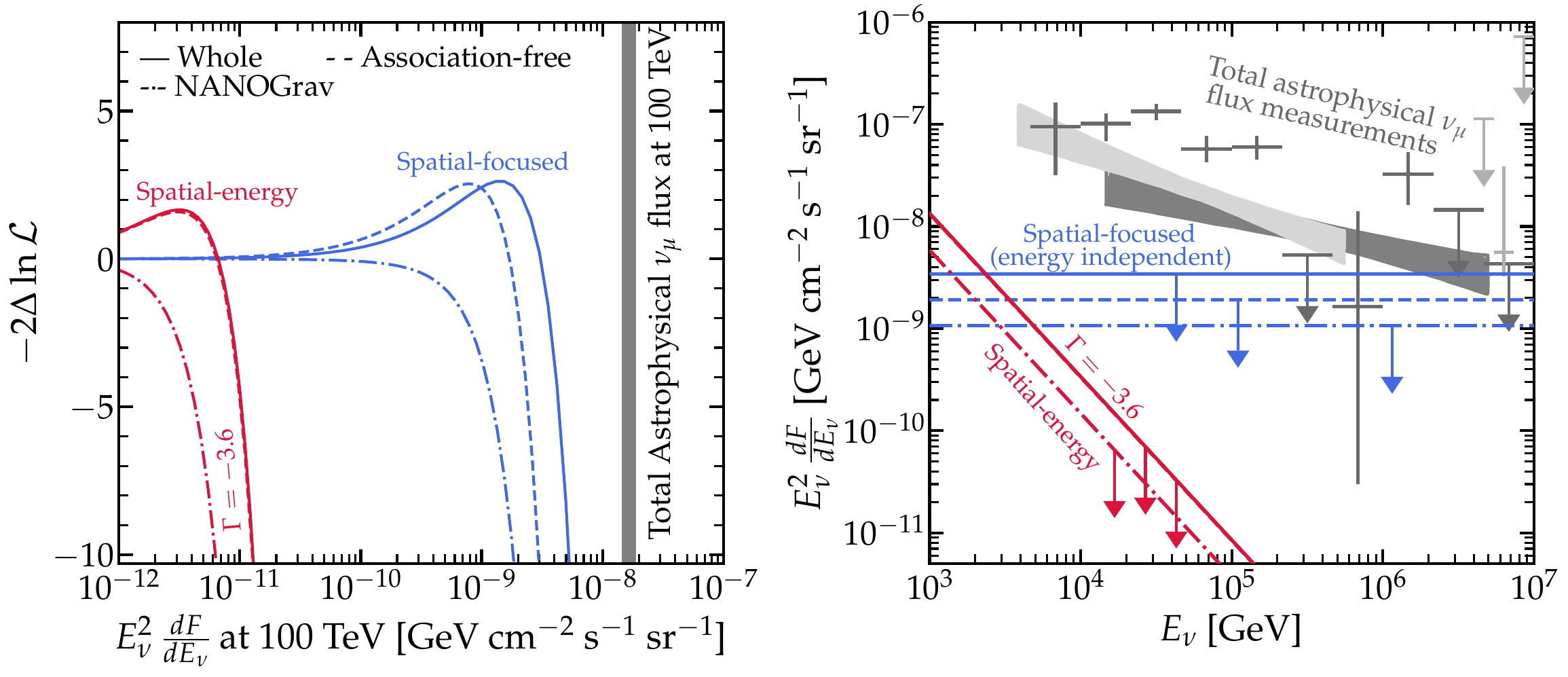}
\caption{
Same as Fig.~\ref{fig:neutrino_flux_plot} but (1) use the \textbf{bolometric-flux weighting scheme} and (2) 
\textbf{\textcolor[HTML]{DC143C}{red}} and \textbf{\textcolor[HTML]{4159E1}{blue}} lines in the right panel are flux ULs instead of measurements. The spatial-energy analysis results coincide for the whole and association-free catalogs in both panels. }
\label{fig:lbol/dl^2_plot}
\end{figure*}

\begin{figure*}[htbp!]
\centering
\includegraphics[width=1\textwidth]{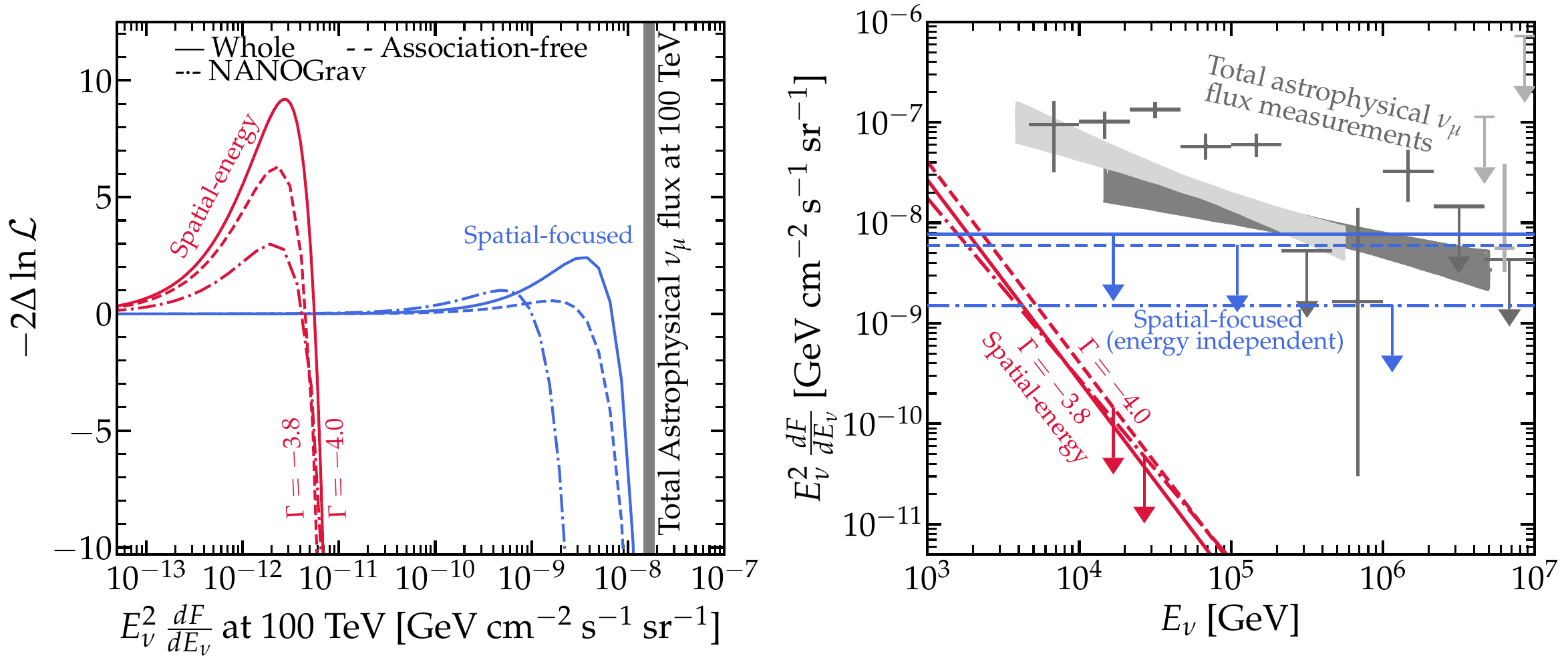}
\caption
{Same as Fig.~\ref{fig:neutrino_flux_plot} but 
(1) use the \textbf{uniform weighting scheme}, 
(2) 
all the curves are flux ULs instead of measurements, except for the red solid,
(3) numbers of sources are different, see table~\ref{tab_results}.}
\label{fig:uniform_plot}
\end{figure*}

Fig.~\ref{fig:neutrino_flux_plot} and table~\ref{tab_results} show our results of this weighting scheme. 

\textbf{Spatial-focused analysis}:
our analysis yields a signal with significance of \textbf{{\color{magenta}$\gtrsim$\,3$\sigma$, 2.9$\sigma$, and 2.8$\sigma$}} (TS = 9.09, 8.35, and 7.59) from our ``whole catalog'', ``association-free catalog'', and the ``NANOGrav catalog'', respectively.

\textbf{Spatial-energy analysis}: 
our analysis yields a signal with significance of \textbf{{\color{magenta}$\gtrsim$\,2.5$\sigma$}}, 2.3$\sigma$, and 2.0$\sigma$ (TS = 10.27, 9.28, and 7.78) at $\Gamma = -3.35$, 3.5, and 3.6 from our ``whole catalog'', ``association-free catalog'', and the ``NANOGrav catalog'', respectively. 
The lower significance here, despite higher TS, compared to those in the spatial-focused analysis, is because of (1) one more free parameter in the analysis and (2) the underlying spectrum possibly being different from a power law.

\medskip

We find that the top three candidates from our default ``whole catalog'' that drive the significance are SDSS J140600.26+013252.2, J153705.95+005522.8, and J0814+0602. 
We sequentially remove these sources from our ``whole catalog'' in the spatial-focused analysis. 
First, removing SDSS J140600.26+013252.2 reduces the significance to 1.9$\sigma$ (TS = 3.53). 
Then, further removing J153705.95+005522 lowers the TS to 1$\sigma$ (TS = 1.03). 
Next, further removing J0814+0602 decreases the significance and TS to zero. These three sources are indicated by  \textcolor[HTML]{CC00CC}{\textbf{encircled pink markers}} in Fig.~\ref{fig:source_loc}. Notably, SMBHB candidate SDSS J140600.26+013252.2 is also included in the targeted GW search by the NANOGrav Collaboration~\cite{Agarwal:2025cag}. This source also exhibits significant $\gamma$-ray emission in the 100 MeV--500 GeV range, with a detection significance of $\sim4.8\sigma$, making it a promising source worth investigating further for cosmic ray emission. More details can be found in Section\,\ref{sec:appendix_gamma}.

\begin{figure}[htbp!]
\centering
\includegraphics[width=1\columnwidth]{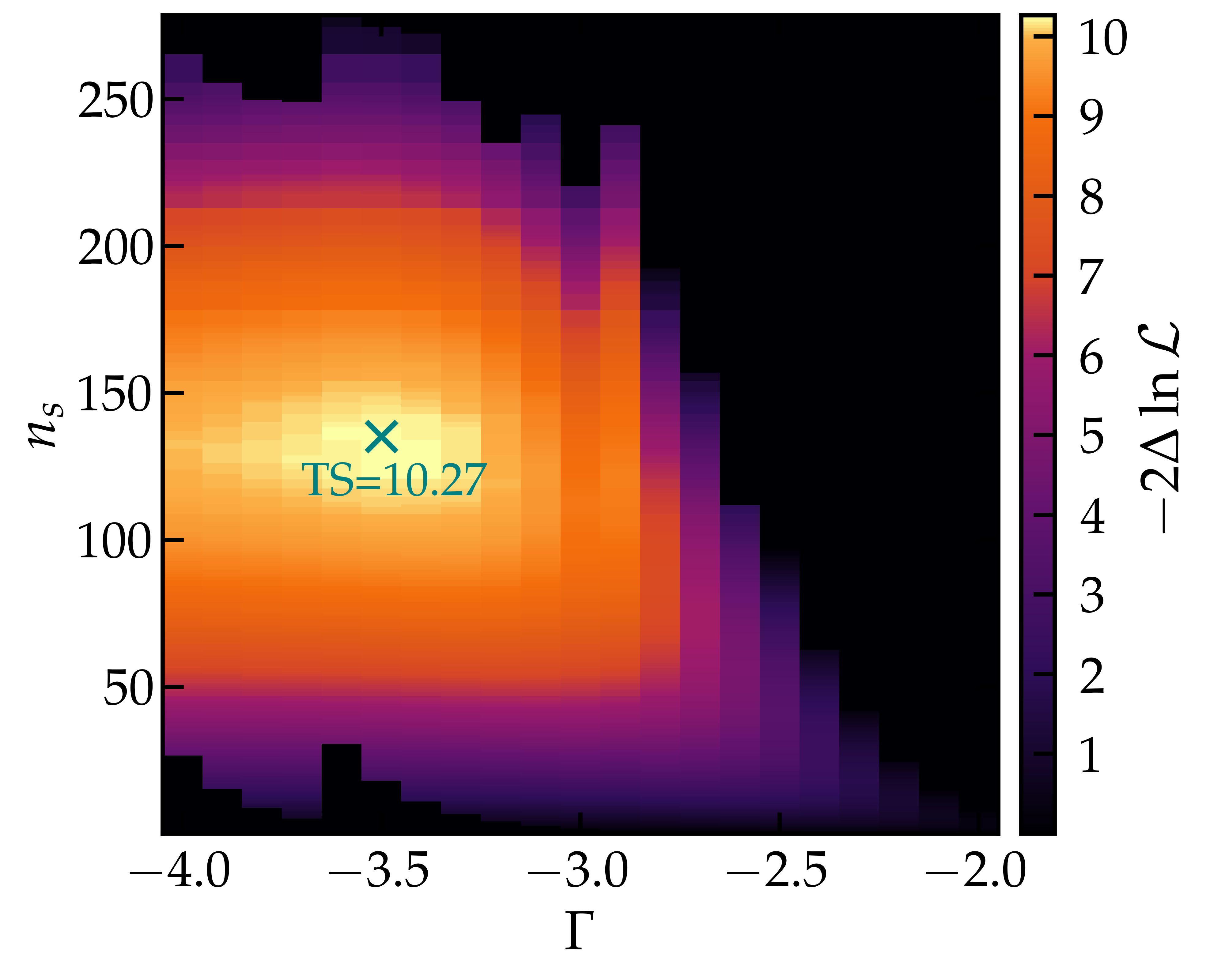}    
\caption{Heatmap of $-2\Delta\ln\mathcal{L}$ as a function of the spectral index $\Gamma$ and the best-fit number of signal events $n_s$ for the case of neutrino-flux weighting scheme and our ``whole catalog.'' 
A custom color scale is adopted to improve visual contrast across the parameter space. The \textbf{\textcolor[HTML]{008080}{teal cross}} indicates the parameter values that maximize the TS.}
\label{fig:TS_Heatmap}
\end{figure}

In Fig.~\ref{fig:TS_Heatmap}, we show the heatmap of $-2 \Delta\rm{ln}\mathcal{L}$ in the $\Gamma$-$n_s$ plane for the neutrino-flux weighting scheme using our default ``whole catalog''. 
The $-2 \Delta\rm{ln}\mathcal{L}$ values remain relatively low (below $\sim 4$) for $\Gamma \gtrsim -2.5$, but increase steadily as $\Gamma$ decreases, reaching TS = 10.27 at $\Gamma = -3.5$ and $\hat{n}_s \approx 135.19$. 
This behavior suggests that the likelihood analysis is driven by low-energy neutrino events, similar to the case of NGC 1068~\cite{IceCube:2019cia, IceCube:2022der} and others.

\subsection{Bolometric-flux weighting scheme}

Fig.~\ref{fig:lbol/dl^2_plot} and table~\ref{tab_results} show our results of this weighting scheme.

\textbf{Spatial-focused analysis}: The highest significance is obtained from our ``whole catalog'' and ``association-free catalog'', which is about 1.6$\sigma$. 
No significance is found from the ``NANOGrav catalog''.

\textbf{Spatial-energy analysis}: our analysis yields no significant signal from any catalog. TS values for all three catalogs are small across our considered $\Gamma$ range, with a maximum of $\rm TS = 1.66$ (0.2$\sigma$) observed from our ``whole catalog'' at $\Gamma=-3.6$.

\subsection{Uniform weighting scheme}

Fig.~\ref{fig:uniform_plot} and table~\ref{tab_results} show our results of this weighting scheme. 

\textbf{Spatial-focused analysis}: No catalog yields a significant signal. We find the highest significance of 1.6$\sigma$ for our ``whole catalog''.
Our ``association-free catalog'' and the ``NANOGrav catalog'' yield 0.74$\sigma$ and 1$\sigma$, respectively.

\textbf{Spatial-energy analysis}:  Our ``whole catalog'' yields a significance of 2.3$\sigma$ (TS = 9.19) at $\Gamma=-4$. We have verified that the TS does not increase for softer spectra. 
The other two catalogs yield smaller significance, with 1.7$\sigma$ for our ``association-free catalog'' and 0.7$\sigma$ for the ``NANOGrav catalog''.

\bigskip
In summary, our results in the neutrino-flux weighting scheme hint at a possible neutrino signal. 
The signal is not influenced by removing sources in our catalog that are close to known neutrino hotspots, as shown by comparing the results from our ``associate-free catalog'' and ``whole catalog''. The significance we find is overall higher for more physically motivated weighting schemes.

Note that the sources we use in our analysis do not form a complete catalog due to telescope sensitivity limits, irregular sky coverage, and the long observational times required to confirm SMBHB signatures. 
Many SMBHBs in the Universe thus remain undetected. 
Ideally, one would correct for this by introducing a completeness factor that accounts for the true underlying population. 
However, a robust estimate of this factor is extremely challenging and requires knowledge of the Eddington ratio across different masses and redshifts, along with several additional ingredients, such as bolometric corrections, duty cycles, and merger-rate evolution. 
At present, no well-established or empirically calibrated distribution of Eddington ratios for SMBHBs exists. Without this, the uncertainties in any completeness estimate become very large and dominated by model assumptions rather than data. As a result, we leave this for future work.

\section{Fraction of energy taken away by HE neutrinos}
\label{sec:gw_smbhb}
\begin{figure}[htbp!]
\centering
\includegraphics[width=1\columnwidth]{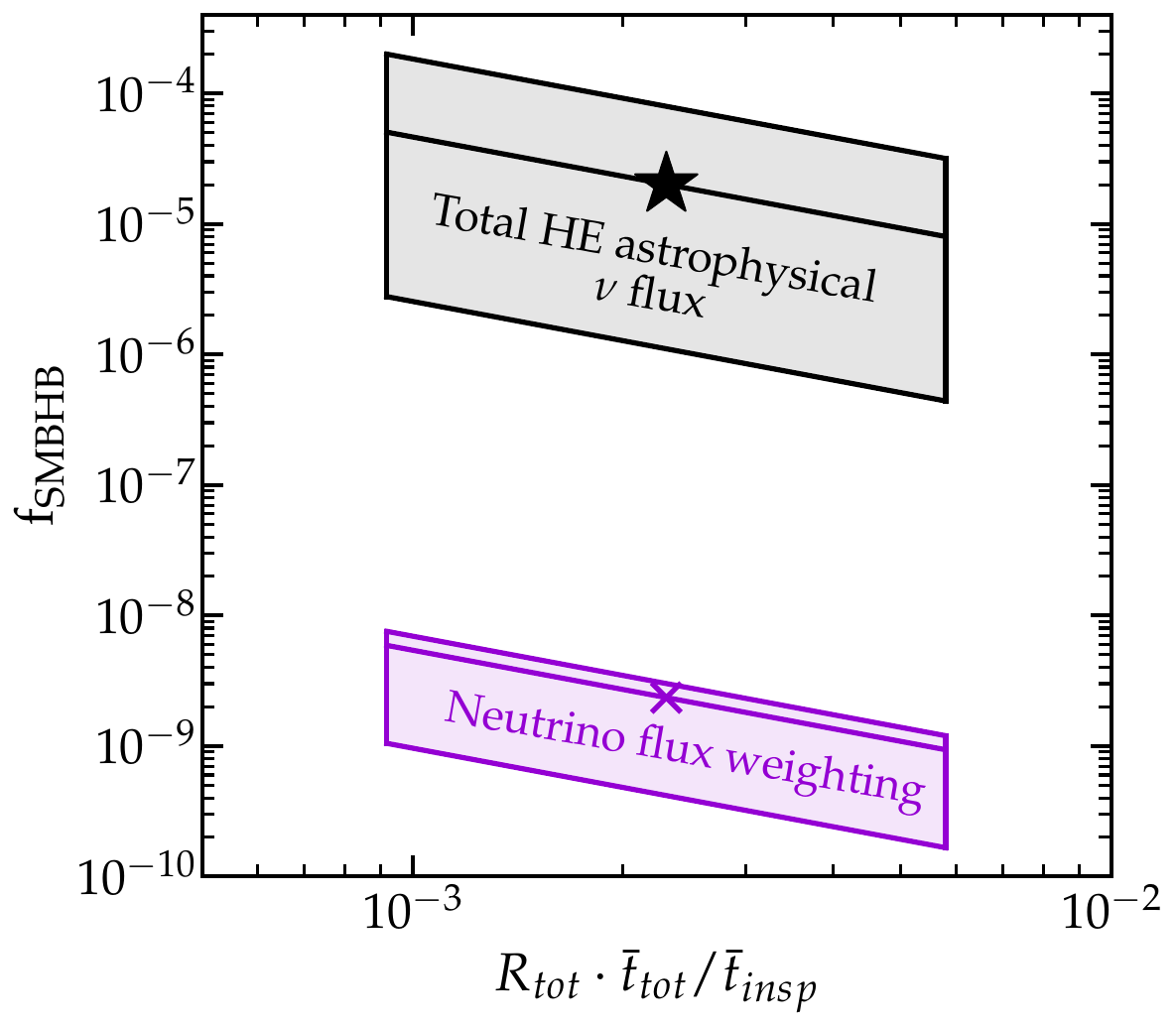}
\caption{Ratio of energies taken away by HE neutrinos, converted from our inferred neutrino flux at 100 TeV, to those by GWs vs. the scaled SMBHB merger rate. 
The \textcolor[HTML]{9400D3}{\textbf{purple cross}} is obtained from converting the inferred flux from the {{neutrino-flux}} weighting schemes that give the highest-significant signal (3$\sigma$), and the uncertainty band covers the region corresponding to the UL. 
As a comparison, the \textbf{black star} denotes the $f_{\rm SMBHB}$ converted from the total HE astrophysical neutrino flux~\cite{Jaroschewski_2022}. 
}\label{fig:f_SMBHBS}
\end{figure}

An interesting question is estimating the fraction of energy taken away by HE neutrinos compared to the total, which is basically the GW energy. 
We use Eq. (4.5) from Ref.~\cite{Jaroschewski_2022} (arXiv version)
to quantify the fraction
\begin{equation}
f^{\nu}_{\text{SMBHB}} = \frac{E_{\nu}^\Gamma \, \Phi(E_{\nu})|_{\text{obs}}}{\kappa_{\Gamma} \, c \, t_H \,h(q) \, \xi_z} \, \frac{1}{M_0 \, c^2} \, \frac{1}{R_{tot}}\,\frac{\bar{t}_{insp}}{\bar{t}_{tot}}\,,
\label{eq:fSMBHB}
\end{equation}

where $E_{\nu}^\Gamma\Phi(E_{\nu})|_{\mathrm{obs}}$ represents the observed neutrino energy flux at Earth. The $h(q)$ term denotes the percentage of the SMBH mass emitted into GWs, which depends on the mass ratio $q$ of two SMBHBs. 
Here $\kappa_\Gamma$ is a normalization constant for the assumed $E_\nu^\Gamma$ power law, 
$c$ is the speed of light, 
$t_H$ is the Hubble time, 
and $\xi_z=\int f(z)/(1+z)^2 \,E(z) dz$. 
$R_{\text{tot}}$ is the total SMBHB merger rate accounting for all relevant timescales between two successive mergers, 
$\bar{t}_{\text{tot}}$ is the mean total timescale of an SMBHB merger (including the preceding galaxy merger), 
and $\bar{t}_{insp}$ is the mean duration of the inspiral phase. 
The ratio $\bar{t}_{tot} / \bar{t}_{insp}$ effectively replaces the total merger timescale with the inspiral timescale, as both GW and neutrino emissions occur primarily during the inspiral phase. The term $M_0$ is a characteristic mass. For details, readers are referred to Section 4 of Ref.~\cite{Jaroschewski_2022}.

Fig.~\ref{fig:f_SMBHBS} shows the $f_{\rm SMBHB}$ as a function of the scaled merger rate, $R_{tot} \, \bar{t}_{tot} / \bar{t}_{insp}$. As our catalog is incomplete, the fraction contributed by the considered sources puts a lower bound on the fraction of total energy taken away.

\section{Conclusions}
\label{sec:conc}

Identifying the sources of high-energy astrophysical neutrinos is a critical step in making them much sharper tools for studying both astrophysics and particle physics. 
Despite extensive searches for more than a decade, which revealed several potential individual sources and only one potential source class---X-ray-bright Seyfert galaxies, the origins of these neutrinos remain largely unresolved. Thus, more source classes should be investigated.

In this work, we conduct the first search for HE neutrino emission from supermassive black hole binaries.
SMBHBs are theoretically motivated cosmic particle accelerators capable of producing HE neutrinos, primarily through hadronic interactions between accelerated particles in relativistic jets and dense circum-binary environment. 
Combining the above with their prevalence in the Universe, SMBHBs are a physically well-motivated yet unexplored source class for HE neutrino searches.

We construct a catalog of 693 SMBHB candidates, together with 10 years of IceCube public data, to search for their spatial and spatial-energy correlations. 
We use an unbinned maximum-likelihood-ratio method, a well-established technique in neutrino astrophysics. 
We conduct two types of stacking analyses: spatial-focused and spatial-energy. The former focuses on the spatial clustering correlation between neutrinos and the sources in our catalog. The latter investigates both spatial and spectral correlations.

Our analysis gives positive results. 
The strongest correlations come from our most physically motivated weighting scheme---neutrino-flux weighting, which shows a \textbf{{\color{magenta}3.0$\sigma$}} or \textbf{{\color{magenta}near-3$\sigma$}} significance in our spatial-focused analysis and 2.0--2.5$\sigma$ from our spatial-energy analysis.
The slightly lower significance of the latter analysis, despite higher TS values, is due to 1) the statistical penalty of having one more free parameter and 2) the underlying spectrum possibly being different from a power law.
Other weight schemes, which are less physically motivated, show weaker correlations.
Finally, we derive the ratio of the energies taken away by HE neutrinos to those by GWs, potentially connecting HE neutrinos with GW signals.

Our results provide the \textit{first evidence of SMBHBs being HE neutrino emitters}. 
Eventually establishing this will promote SMBHBs as multimessenger sources, as they are the main targets of pulsar timing arrays for nano-Hz GWs.

High-energy neutrino astrophysics and its essential role in multimessenger astrophysics are becoming increasingly exciting thanks to rapidly expanding observational efforts and new telescopes, motivated by their promises and several major discoveries over the past decade. 
IceCube has been operating for more than 17 years in good condition and will continue. 
KM3NeT~\cite{KM3Net:2016zxf} and Baikal-GVD~\cite{Allakhverdyan:2021vkk, Aynutdinov:2023ifk}, both currently operating and nearing completion to a 1-km$^3$ volume comparable to IceCube, will provide complementary sensitivity in the southern sky compared to IceCube.
The next-generation telescopes, including IceCube-Gen2~\cite{IceCube-Gen2:2020qha}, 
TRIDENT~\cite{TRIDENT:2022hql},
P-ONE~\cite{P-ONE:2020ljt},
NEON~\cite{Zhang:2024slv}, and HUNT~\cite{Huang:2023mzt}, and are envisioned to have instrumented volumes of $\sim 8$–30 km$^3$. 
We hope that the current and next-generation HE neutrino observatories will conclusively test the hypothesis of SMBHBs being potential HE neutrino emitters.

\section*{Acknowledgements}

We thank Anirban Das, Deep Jyoti Das, Jaya Doliya, Achamveedu Gopakumar and Lu Lu for their helpful comments and suggestions. We especially thank Avinash Kumar Paladi for his help in constructing the SMBHB catalog. P.A.D acknowledges financial support from INSPIRE–SHE Fellowship of the Department of Science and Technology, Government of India. S.B. acknowledges the Council of Scientific and Industrial Research (CSIR), Government of India, for supporting his research under the CSIR Junior/Senior Research Fellowship program through grant no. 09/0079(15488)/2022-EMR-I. B.Z. is supported by Fermi Forward Discovery Group, LLC under Contract No. 89243024CSC000002 with the U.S. Department of Energy, Office of Science, Office of High Energy Physics.  R.L. acknowledges financial support from the institute start-up funds and ISRO-IISc STC for the grant no. ISTC/PHY/RL/499.

\paragraph*{Software\,:} Python~\cite{10.5555/1593511}, NumPy~\cite{harris2020array}, SciPy~\cite{2020SciPy-NMeth}, Matplotlib~\cite{Hunter:2007}, Astropy~\cite{price2018astropy}, PyAstronomy~\cite{pya}, multiprocessing~\cite{2012arXiv1202.1056M}, iMinuit~\cite{iminuit,James:1975dr}

\onecolumngrid
\clearpage
\appendix

\setcounter{figure}{0}
\setcounter{table}{0}
\setcounter{equation}{0}

\renewcommand{\thefigure}{A\arabic{figure}}
\renewcommand{\thetable}{A\arabic{table}}
\renewcommand{\theequation}{A\arabic{equation}}

\section{Gamma-ray detection for the leading HE-neutrino-emitting SMBHB candidates}
\label{sec:appendix_gamma}

\begin{figure*}[!htbp]  
\centering
\includegraphics[width=0.48\textwidth]{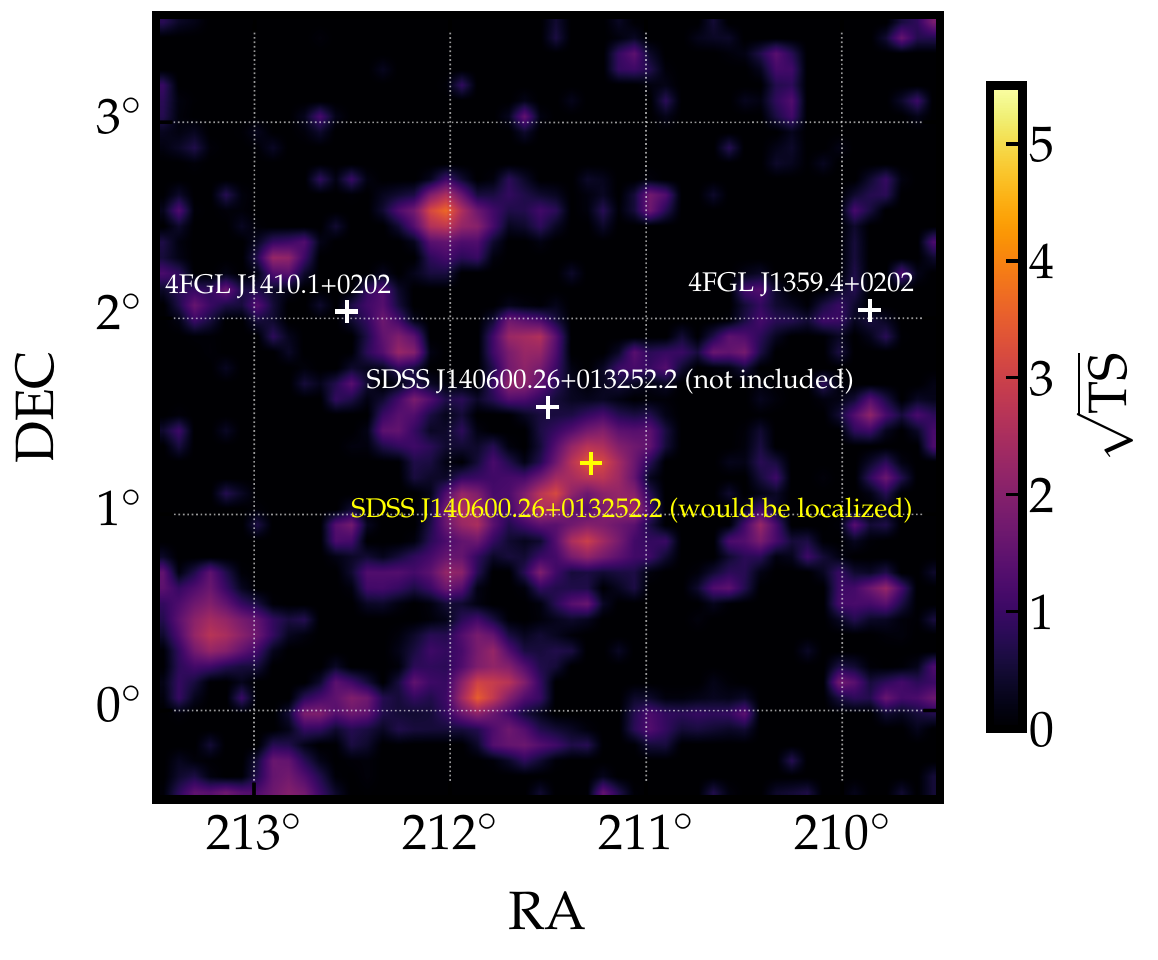}
\hfill
\includegraphics[width=0.48\textwidth]{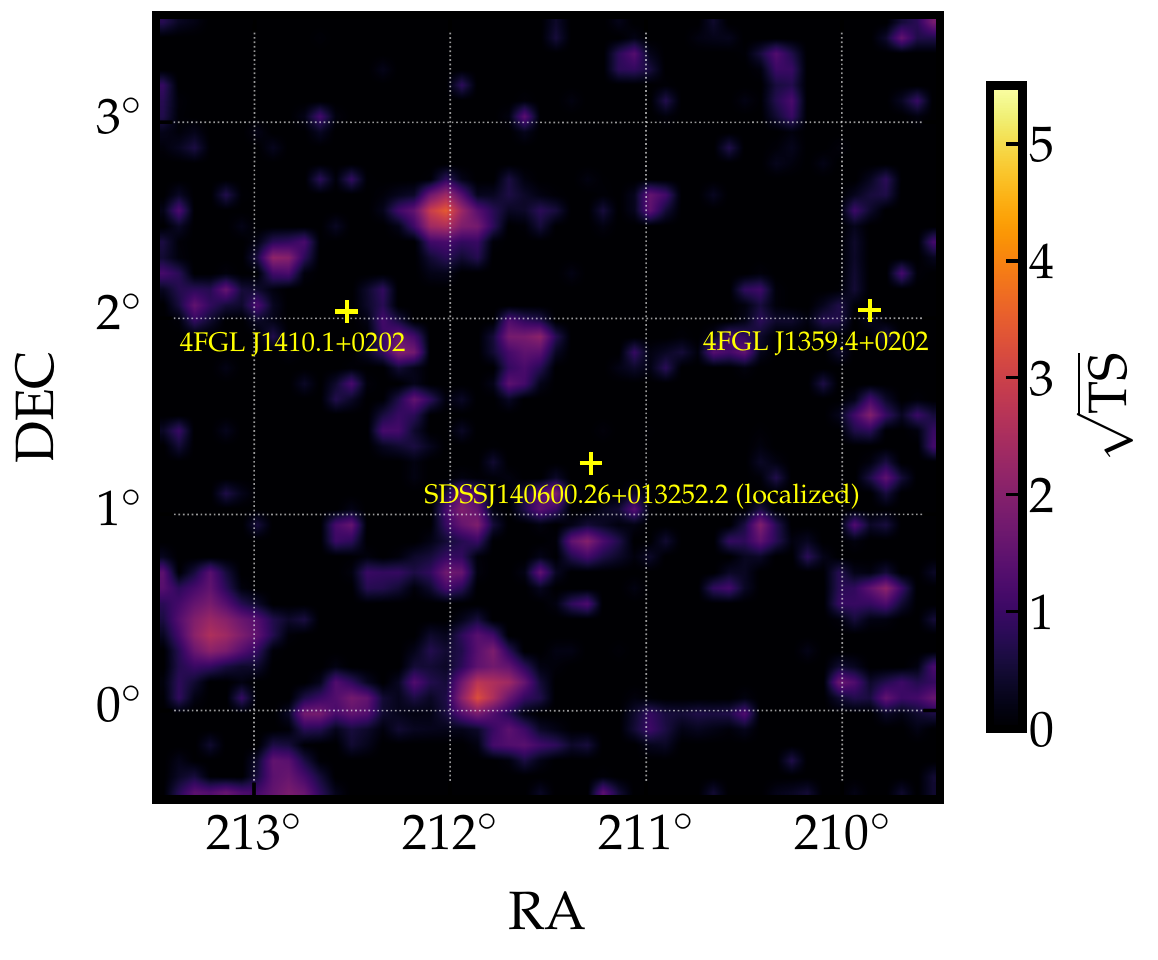}
\caption{
Gamma-ray $\rm \sqrt{TS}$ ($\simeq$ significance) residual maps from \textit{Fermi}-LAT data (0.1 GeV–500 GeV) centered on the optimized position of SDSS J140600.26+013252.2, the most dominant source in our neutrino-flux weighting scheme. Background sources from the 14-year {\it Fermi}-LAT catalog are also shown. \textbf{Left}: residual map without including SDSS J140600.26+013252.2 in the analysis (but shown in the figure as a reference). 
White markers indicate catalog coordinates, while the yellow markers show the coordinates from our analysis. \textbf{Right}: residual map after including SDSS J140600.26+013252.2 into our analysis.
}
\label{fig:ts_map}
\end{figure*}

\begin{table*}[ht]
\centering
\renewcommand{\arraystretch}{1.1}
\vspace{0.3em}
{\footnotesize
\setlength{\tabcolsep}{5pt}
\renewcommand{\arraystretch}{1.5}
\sisetup{separate-uncertainty=true, table-align-text-post=false}
\begin{tabular}{
    l|c|c|c|c|c|c
}
\hline\hline
\textbf{Name} 
& \textbf{RA, Dec} 
& \textbf{RA, Dec ($\gamma$)} 
& \textbf{Spec. index} 
& \textbf{Photon flux} 
& \textbf{Energy flux} 
& \textbf{TS, sig.} \\
 & {[deg]} & {[deg]} 
 & {($\Gamma_\gamma$)} 
 & {[cm$^{-2}$\,s$^{-1}$]} 
 & {[GeV\,cm$^{-2}$s$^{-1}$]} 
 & \\
\hline

SDSS J140600.26+013252.2 
& 211.50,\,1.54
& 211.28,\,1.26
& -2.67\,$\pm$\,0.21
& (3.69\,$\pm$\,1.27)$\times10^{-9}$ 
& (9.14\,$\pm$\,2.26)$\times10^{-10}$ 
& 23.1, \textbf{\textcolor{teal}{4.8$\sigma$}} \\
\hline

J153705.95+005522.8 
& 234.27,\,0.92
& 234.15,\,0.67
&-2.04\,$\pm$\,0.35
& (5.78\,$\pm$\,5.73)$\times10^{-10}$ 
& (4.35\,$\pm$\,4.61)$\times10^{-10}$ 
& 5.98, 2.5$\sigma$ \\

\hline

J0814+0602 
& 123.52,\,6.04 
& 123.75,\,6.22
& -1.92\,$\pm$\,0.16
& (1.86\,$\pm$\,1.22)$\times10^{-10}$ 
& (2.08\,$\pm$\,1.52)$\times10^{-10}$ 
& 5.02, 2.2$\sigma$ \\

\hline\hline

\end{tabular}}

\caption{Summary of the {\it Fermi}-LAT gamma-ray analysis for the three dominant sources in the neutrino-flux weighting scheme. The columns from left to right are: source name, cataloged coordinates, optimized coordinates, spectral index, integrated photon flux, energy flux, and TS value. The fluxes are integrated over the energy range of $0.1\text{ GeV}$ to $500\text{ GeV}$.}

\label{tab:FR0s}
\end{table*}

In this section, we search for the gamma-ray emission using \textit{Fermi}-LAT data from the three sources that contribute the most to our $3\sigma$ significance in the HE-neutrino search (Sec.~\ref{sec_results_nuflux}).
As both gamma-rays and neutrinos are expected to originate from the same hadronic processes, 
gamma-ray detection would provide additional support for the sources being HE neutrino emitters and, hence, HE cosmic-ray emitters.

\paragraph*{\textbf{Methodology:}}
For each source, we perform a standard binned-likelihood analysis using \texttt{Fermipy} (v1.3.1)\,\cite{2017ICRC...35..824W} built on \texttt{Fermitools} (v2.2.0). 
We use Pass~8 (P8R3 SOURCE, optimized for point-source analysis) data in the 0.1~GeV--500~GeV energy range, applying the recommended quality cuts (\texttt{DATA\_QUAL>0}, \texttt{LAT\_CONFIG==1}) and a zenith-angle cut of $90^\circ$ to remove contamination from the Earth limb. 
Each region of interest (ROI) is defined as a $15^\circ \times 15^\circ$ WCS grid with a spatial bin size of $0.08^\circ$ and eight logarithmic energy bins per decade. The background model includes all 4FGL-DR4 sources within a $20^\circ \times 20^\circ$ region, the Galactic diffuse template \texttt{gll\_iem\_v07}, and the isotropic component \texttt{iso\_P8R3\_SOURCE\_V3\_v1}. 
The candidate sources are modeled as point emitters with a simple power law, $dN_\gamma/dE_\gamma = N_{\gamma,0} \left( {E_\gamma}/{E_{\gamma, 0}} \right)^{\Gamma_\gamma},
$ with both $N_{\gamma,0}$ and $\Gamma_\gamma$ treated as free parameters. Detection significance is quantified using the test statistic ${\rm TS} = 2\ln(L_{\rm src,max}/L_{\rm null,max})$, where $L_{\rm src,max}$ is the maximized likelihood including a new source in the model and $L_{\rm null,max}$ is the maximized likelihood in the absence of it. For each ROI, we iteratively generate TS maps with \texttt{gta.find\_source} to identify unresolved excesses, update the source model, and finally perform a full maximum-likelihood optimization.

\paragraph*{\textbf{Results}:}
Table~\ref{tab:FR0s} summarizes our main results from the gamma-ray analysis.
We detected gamma rays with \textbf{\textcolor{teal}{4.8$\sigma$}} ($\rm TS = 23.10$) from our most dominant source, SDSS~J140600.26+013252.2. For the next two dominant sources, we detect mild gamma-ray emissoin, with 2.5$\sigma$ and 2.2$\sigma$ ($\rm TS = 5.98$ and $5.02$), respectively.

Fig.~\ref{fig:ts_map} shows the TS map for a $4^\circ \times 4^\circ$ region of interest centered on SDSS~J140600.26+013252.2.
Comparing the two panels, adding a point source at the location of SDSS~J140600.26+013252.2 explains the observed excess. As a result, SDSS SDSS~J140600.26+013252.2 is detected with a TS of 23.10, corresponding to a significance of $\sim 4.8\sigma$. The gamma-ray emission is modeled by a power-law spectrum given by $dN_\gamma/dE_\gamma = (1.31 \pm 0.34) \times 10^{-10} \left( E_\gamma/1\,\text{GeV} \right)^{-(2.67 \pm 0.21)} \, \text{ph}\,\text{cm}^{-2}\,\text{s}^{-1}\,\text{GeV}^{-1}$.

\setcounter{figure}{0}
\setcounter{table}{0}
\setcounter{equation}{0}

\renewcommand{\thefigure}{B\arabic{figure}}
\renewcommand{\thetable}{B\arabic{table}}
\renewcommand{\theequation}{B\arabic{equation}}

\section{ Broken Power-Law Spectrum Does Not Change Our Results}
\label{sec:appendix_bpl}
\begin{figure*}[!htbp]
\centering
\includegraphics[width=0.5\textwidth]{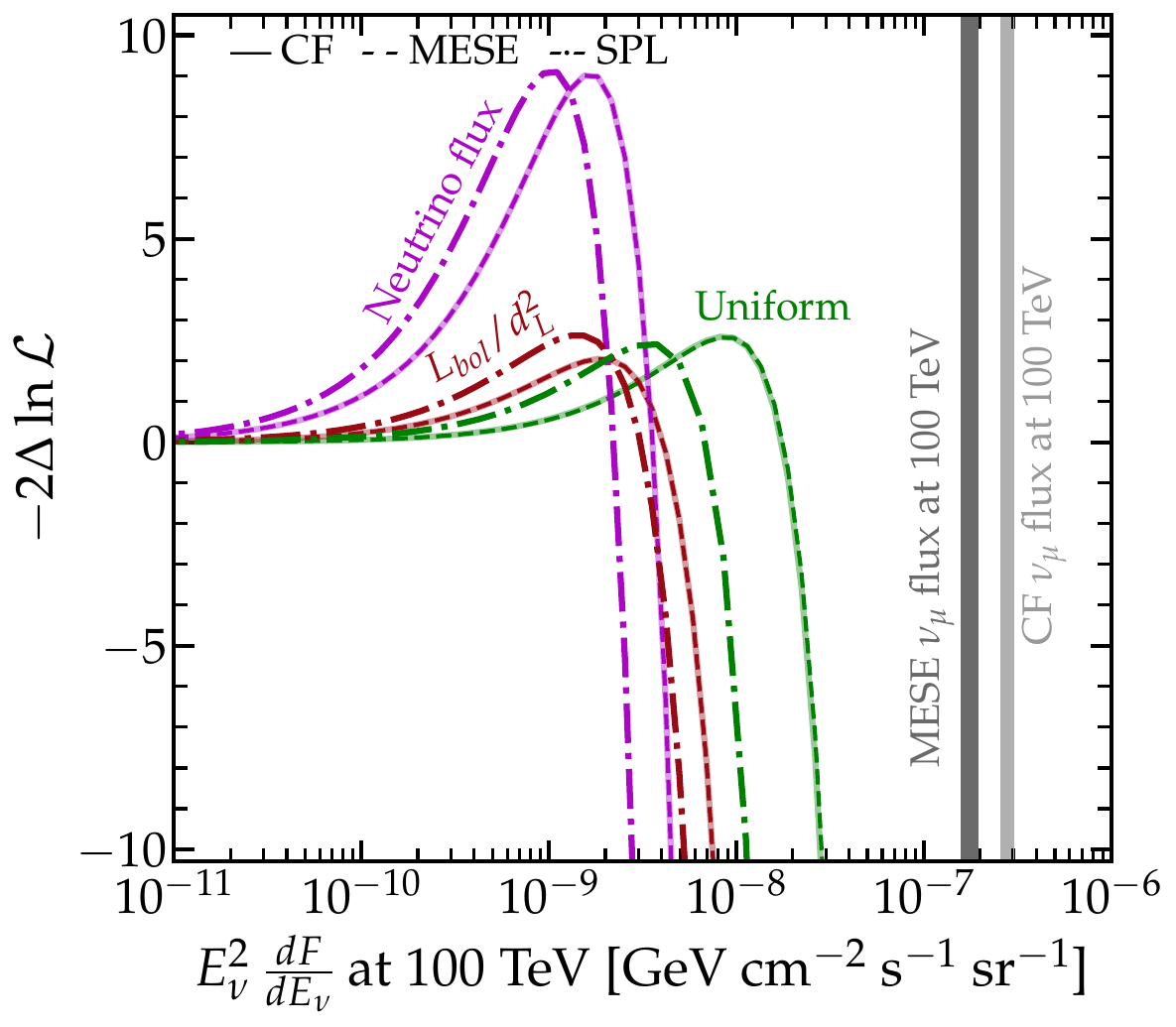}
\caption{Log-likelihood values as a function of $E_\nu^2 dF/dE_\nu$ at 100~TeV. Stacking-analysis results for our default ``whole catalog'' using single-power-law spectrum and two broken-power-law spectral models—the Combined Fit (CF) and the Medium Energy Starting Events (MESE) models~\cite{IceCube:2025tgp,IceCube:2025ewu}—are shown for the spatial-focused analysis. Results are presented for all three weighting schemes: \textbf{\textcolor[HTML]{A906C6}{neutrino-flux (violet)}}, \textbf{\textcolor[HTML]{990B14}{bolometric-flux (maroon; \(L_{bol}/d_L^2\))}}, and \textbf{\textcolor[HTML]{008000}{uniform (green)}}. Solid curves correspond to the CF spectral index, while dashed curves correspond to the MESE spectral index; for visual clarity, CF curves are displayed with slightly reduced opacity relative to MESE. The MESE and CF results largely overlap. Dash-dotted curves show the corresponding single-power-law results, identical to those presented in Sec.~\ref{sec_results}. The shaded vertical bands indicate IceCube measurements of the diffuse astrophysical neutrino flux at $100$~TeV: the \textcolor{darkgray}{\textbf{CF}} determination from the combined track+cascade analysis (\textcolor[HTML]{b0b0b0}{$\blacksquare$}) and the \textcolor{black}{\textbf{MESE}} determination (\textcolor[HTML]{6a6a6a}{$\blacksquare$}).
}
\label{fig:Stacking_broken_power_law_MESE_CF} 
\end{figure*}

A recent analysis by the IceCube Collaboration~\cite{IceCube:2025tgp, IceCube:2025ewu} found that the HE astrophysical neutrino spectrum is better fit by a broken-power-law than a single-power-law with a significance of $\sim 4.5\sigma$. We assume that the SMBHB candidates follow a broken power-law neutrino spectrum with best-fit spectral indices and break energy taken from the Combined Fit (CF) and Medium Energy Starting Events (MESE) models\,\cite{IceCube:2025tgp, IceCube:2025ewu}. We conduct the stacking analysis following the procedure mentioned in Sec.~\ref{sec_formalism_stacking}.

Figure~\ref{fig:Stacking_broken_power_law_MESE_CF} presents the spatial-focused stacking-analysis results for the single-power-law and the two broken-power-law spectral models. We find that all three weighting schemes yield TS values that are essentially unchanged when moving from the single-power-law to the broken-power-law parametrizations, reflecting the fact that our energy-integrated (spatial-focused) search is only weakly sensitive to the assumed spectral shape. The broken-power-law fits, however, prefer slightly higher flux normalizations ($E_\nu^2\,dF/dE_\nu$ at 100~TeV) compared to the single-power-law model, indicating that a larger overall astrophysical neutrino flux is required to reproduce the IceCube data. Since the TS values remain stable across single-power-law and broken-power-law models, we conclude that our stacking-analysis results are robust and remain unchanged under the broken-power-law HE neutrino spectrum.

\setcounter{figure}{0}
\setcounter{table}{0}
\setcounter{equation}{0}

\renewcommand{\thefigure}{E\arabic{figure}}
\renewcommand{\thetable}{E\arabic{table}}
\renewcommand{\theequation}{E\arabic{equation}}

\twocolumngrid

\bibliography{ref.bib}

\end{document}